\documentclass[twocolumn]{revtex4}
\usepackage{amsmath}

\begin{document}

\title{Optical transformation from chirplet to fractional Fourier
transformation kernel }
\author{Hong-yi Fan and Li-yun Hu*}
\affiliation{Department of Physics, Shanghai Jiao Tong University, Shanghai, 200030, P.R.
China\\
Corresponding author. hlyun2008@126.com or hlyun@sjtu.edu.cn}

\begin{abstract}
{\small We find a new integration transformation which can convert a
chirplet function to fractional Fourier transformation kernel, this new
transformation is invertible and obeys Parseval theorem. Under this
transformation a new relationship between a phase space function and its
Weyl-Wigner quantum correspondence operator is revealed.}
\end{abstract}

\maketitle

In the history of developing optics we have known that each optical setup
corresponds to an optical transformation, for example, thick lens as a
fractional Fourier transformer. In turn, once a new integration transform is
found, its experimental implementation is expected, for example, the
fractional Fourier transform (FrFT) of a function was originally introduced
by Namias as a mathematical tool for solving theoretical physical problems
\cite{1,2}, and later Mendlovic, Ozakatas et al explored its applications in
optics by redefining it as the change of the field caused by propagation
along a quadratic Graded-Index (GRIN) medium\cite{3,4,5,6,7,8,9,10,11}. In
this Letter we report a new integration transformation which can convert
chirplet function to fractional Fourier transformation kernel, as this new
transformation is invertible and obeys Parseval theorem, we expect it be
realized by experimentalists.

The new transform we propose here is
\begin{equation}
\iint_{-\infty }^{\infty }\frac{dpdq}{\pi }e^{2i\left( p-x\right) \left(
q-y\right) }h(p,q)\equiv f\left( x,y\right) ,  \label{1}
\end{equation}%
which differs from the usual two-fold Fourier transformation $\iint_{-\infty
}^{\infty }\frac{dxdy}{4\pi ^{2}}e^{ipx+iqy}f(x,y).$ In particular, when $%
h(p,q)=1,$ Eq. (\ref{1}) reduces to
\begin{equation}
\iint_{-\infty }^{\infty }\frac{dpdq}{\pi }e^{2i\left( p-x\right) \left(
q-y\right) }=\int_{-\infty }^{\infty }dq\delta \left( q-y\right)
e^{-2xi\left( q-y\right) }=1,  \label{2}
\end{equation}%
so $e^{2i\left( p-x\right) \left( q-y\right) }$ can be considered a basis
funtion in $p-q$ phase space, or Eq. (\ref{1}) can be looked as an expansion
of $f\left( x,y\right) $ with the expansion coefficient being $h(p,q).$ We
can prove that the reciprocal transformation of (\ref{1}) is%
\begin{equation}
\iint_{-\infty }^{\infty }\frac{dxdy}{\pi }e^{-2i(p-x)(q-y)}f(x,y)=h(p,q).
\label{3}
\end{equation}%
In fact, substituting (\ref{1}) into the left-hand side of (\ref{3}) yields%
\begin{eqnarray}
&&\iint_{-\infty }^{\infty }\frac{dp^{\prime }dq^{\prime }}{\pi }h(p^{\prime
},q^{\prime })\iint \frac{dxdy}{\pi }e^{2i\left[ \left( p^{\prime }-x\right)
\left( q^{\prime }-y\right) -\left( p-x\right) \left( q-y\right) \right] }
\notag \\
&=&\iint_{-\infty }^{\infty }dp^{\prime }dq^{\prime }h(p^{\prime },q^{\prime
})e^{2i\left( p^{\prime }q^{\prime }-pq\right) }\delta \left( p-p^{\prime
}\right) \delta \left( q-q^{\prime }\right) =h(p,q).  \label{4}
\end{eqnarray}%
This transformation's Parseval-like theorem is
\begin{eqnarray}
&&\iint_{-\infty }^{\infty }\frac{dpdq}{\pi }|h(p,q)|^{2}  \notag \\
&=&\iint \frac{dxdy}{\pi }|f\left( x,y\right) |^{2}\iint \frac{dx^{\prime
}dy^{\prime }}{\pi }e^{2i\left( x^{\prime }y^{\prime }-xy\right) }  \notag \\
&&\times \iint_{-\infty }^{\infty }\frac{dpdq}{\pi }e^{2i\left[ \left(
-y^{\prime }p-x^{\prime }q\right) +\left( py+xq\right) \right] }  \notag \\
&=&\iint \frac{dxdy}{\pi }|f\left( x,y\right) |^{2}\iint dx^{\prime
}dy^{\prime }e^{2i\left( x^{\prime }y^{\prime }-xy\right) }  \notag \\
&&\times \delta \left( x-x^{\prime }\right) \delta \left( p-p^{\prime
}\right)   \notag \\
&=&\iint \frac{dxdy}{\pi }|f\left( x,y\right) |^{2}.  \label{5}
\end{eqnarray}%
Now we apply Eq. (\ref{1}) to phase space transformation in quantum optics.
Recall that a signal $\psi \left( q\right) $'s Wigner transform \cite%
{12,13,14,15} is
\begin{equation}
\psi \left( q\right) \rightarrow \int \frac{du}{2\pi }e^{ipu}\psi ^{\ast
}\left( q+\frac{u}{2}\right) \psi \left( q-\frac{u}{2}\right) .  \label{6}
\end{equation}%
Using Dirac's symbol \cite{16} to write $\psi \left( q\right) =\left\langle
q\right\vert \left. \psi \right\rangle ,$ $\left\vert q\right\rangle $ is
the eigenvector of coordinate $Q$, $Q\left\vert q\right\rangle =q\left\vert
q\right\rangle ,$ $\left[ Q,P\right] =i\hbar ,$ the Wigner operator emerges
from (\ref{6}),%
\begin{equation}
\frac{1}{2\pi }\int_{-\infty }^{\infty }due^{-ipu}\left\vert q-\frac{u}{2}%
\right\rangle \left\langle q+\frac{u}{2}\right\vert =\Delta \left(
p,q\right) ,\text{ }\hbar =1.  \label{7}
\end{equation}%
If $h\left( q,p\right) $ is quantized as the operator $\hat{H}\left(
P,Q\right) $ through the Weyl-Wigner correspondence \cite{17}%
\begin{equation}
H\left( P,Q\right) =\iint_{-\infty }^{\infty }dpdq\Delta \left( p,q\right)
h\left( q,p\right) ,  \label{8}
\end{equation}%
then%
\begin{equation}
h\left( q,p\right) =\int_{-\infty }^{\infty }due^{-ipu}\left\langle q+\frac{u%
}{2}\right\vert \hat{H}\left( Q,P\right) \left\vert q-\frac{u}{2}%
\right\rangle ,  \label{9}
\end{equation}%
this in the literature is named the Weyl transform, $h\left( q,p\right) $ is
the Weyl classical correspondence of the operator $\hat{H}\left( Q,P\right) $%
. Substituting (\ref{9}) into (\ref{1}) we have
\begin{eqnarray}
&&\iint_{-\infty }^{\infty }\frac{dpdq}{\pi }e^{2i\left( p-x\right) \left(
q-y\right) }h(p,q)  \notag \\
&=&\iint_{-\infty }^{\infty }\frac{dpdq}{\pi }e^{2i\left( p-x\right) \left(
q-y\right) }\int_{-\infty }^{\infty }due^{-ipu}  \notag \\
&&\times \left\langle q+\frac{u}{2}\right\vert \hat{H}\left( Q,P\right)
\left\vert q-\frac{u}{2}\right\rangle   \notag \\
&=&\int_{-\infty }^{\infty }dq\int_{-\infty }^{\infty }du\left\langle q+%
\frac{u}{2}\right\vert \hat{H}\left( Q,P\right) \left\vert q-\frac{u}{2}%
\right\rangle   \notag \\
&&\times \delta \left( q-y-\frac{u}{2}\right) e^{-2ix\left( q-y\right) }
\notag \\
&=&\int_{-\infty }^{\infty }due^{-ixu}\left\langle y+u\right\vert \hat{H}%
\left( Q,P\right) \left\vert y\right\rangle .  \label{10}
\end{eqnarray}%
Using $\left\langle y+u\right\vert =\left\langle u\right\vert e^{iPy}$ and $(%
\sqrt{2\pi })^{-1}e^{-ixu}=\left\langle p_{=x}\right\vert \left.
u\right\rangle ,$ where $\left\langle p\right\vert $ is the momentum
eigenvector, and%
\begin{eqnarray}
\int_{-\infty }^{\infty }due^{-ixu}\left\langle y+u\right\vert
&=&\int_{-\infty }^{\infty }due^{-ixu}\left\langle u\right\vert e^{iPy}
\notag \\
&=&\sqrt{2\pi }\int_{-\infty }^{\infty }du\left\langle p_{=x}\right\vert
\left. u\right\rangle \left\langle u\right\vert e^{iPy}  \notag \\
&=&\sqrt{2\pi }\left\langle p_{=x}\right\vert e^{ixy},  \label{11}
\end{eqnarray}%
then Eq. (\ref{10}) becomes%
\begin{eqnarray}
&&\iint_{-\infty }^{\infty }\frac{dpdq}{\pi }e^{2i\left( p-x\right) \left(
q-y\right) }h(p,q)  \notag \\
&=&\sqrt{2\pi }\left\langle p_{=x}\right\vert \hat{H}\left( Q,P\right)
\left\vert y\right\rangle e^{ixy},  \label{12}
\end{eqnarray}%
thus through the new integration transformation a new relationship between a
phase space function $h(p,q)$ and its Weyl-Wigner correspondence operator $%
\hat{H}\left( Q,P\right) $ is revealed. The inverse of (\ref{12}), according
to (\ref{3}), is%
\begin{equation}
\iint_{-\infty }^{\infty }\frac{dxdy}{\sqrt{\pi /2}}e^{-2i\left( p-x\right)
\left( q-y\right) }\left\langle p_{=x}\right\vert \hat{H}\left( Q,P\right)
\left\vert y\right\rangle e^{ixy}=h(p,q).  \label{13}
\end{equation}%
For example, when $\hat{H}\left( Q,P\right) =e^{f(P^{2}+Q^{2}-1)/2},$ its
classical correspondence is%
\begin{equation}
e^{f\left( P^{2}+Q^{2}-1\right) /2}\rightarrow h(p,q)=\frac{2}{e^{f}+1}\exp
\left\{ 2\frac{e^{f}-1}{e^{f}+1}\left( p^{2}+q^{2}\right) \right\} .
\label{14}
\end{equation}%
Substituting (\ref{14}) into (\ref{12}) we have%
\begin{eqnarray}
&&\frac{2}{e^{f}+1}\iint_{-\infty }^{\infty }\frac{dpdq}{\pi }e^{2i\left(
p-x\right) \left( q-y\right) }\exp \left\{ 2\frac{e^{f}-1}{e^{f}+1}\left(
p^{2}+q^{2}\right) \right\}   \notag \\
&=&\sqrt{2\pi }\left\langle p_{=x}\right\vert e^{f\left(
P^{2}+Q^{2}-1\right) /2}\left\vert y\right\rangle e^{ixy}.  \label{15}
\end{eqnarray}%
Using the Gaussian integration formula
\begin{eqnarray}
&&\iint_{-\infty }^{\infty }\frac{dpdq}{\pi }e^{2i\left( p-x\right) \left(
q-y\right) }e^{-\lambda \left( p^{2}+q^{2}\right) }  \notag \\
&=&\frac{1}{\sqrt{\lambda ^{2}+1}}\exp \left\{ \frac{-\lambda \left(
x^{2}+y^{2}\right) }{\lambda ^{2}+1}+\frac{2i\lambda ^{2}}{\lambda ^{2}+1}%
xy\right\} ,  \label{16}
\end{eqnarray}%
in particular, when%
\begin{equation}
\lambda =-i\tan \left( \frac{\pi }{4}-\frac{\alpha }{2}\right) ,  \label{17}
\end{equation}%
with%
\begin{equation}
\frac{-\lambda }{\lambda ^{2}+1}=\frac{i}{2\tan \alpha },\text{ }\frac{%
2\lambda ^{2}}{\lambda ^{2}+1}=1-\frac{1}{\sin \alpha },  \label{18}
\end{equation}%
Eq. (\ref{16}) becomes$\allowbreak $%
\begin{eqnarray}
&&\frac{2}{ie^{-i\alpha }+1}\iint_{-\infty }^{\infty }\frac{dpdq}{\pi }%
e^{2i\left( p-x\right) \left( q-y\right) }  \notag \\
&&\times \exp \left\{ i\tan \left( \frac{\pi }{4}-\frac{\alpha }{2}\right)
\left( p^{2}+q^{2}\right) \right\}   \notag \\
&=&\frac{1}{\sqrt{i\sin \alpha e^{-i\alpha }}}\exp \left\{ \frac{i\left(
x^{2}+y^{2}\right) }{2\tan \alpha }-\frac{ixy}{\sin \alpha }\right\} e^{ixy},
\label{19}
\end{eqnarray}%
where $\exp \{i\tan \left( \frac{\pi }{4}-\frac{\alpha }{2}\right) \left(
p^{2}+q^{2}\right) \}$ represents an infinite long chirplet function.
Comparing (\ref{19}) with (\ref{15}) we see $ie^{-i\alpha }=e^{f},$ $%
f=i\left( \frac{\pi }{2}-\alpha \right) ,$ it then follows%
\begin{eqnarray}
&&\left\langle p_{=x}\right\vert e^{i(\frac{\pi }{2}-\alpha )\left(
P^{2}+Q^{2}-1\right) /2}\left\vert y\right\rangle   \notag \\
&=&\frac{1}{\sqrt{2\pi i\sin \alpha e^{-i\alpha }}}\exp \left\{ \frac{%
i\left( x^{2}+y^{2}\right) }{2\tan \alpha }-\frac{ixy}{\sin \alpha }\right\}
,  \label{20}
\end{eqnarray}%
where the right-hand side of (\ref{20}) is just the FrFT kernel. Therefore
the new integration transformation (\ref{1}) can convert spherical wave to
FrFT kernel. We expect this transformation could be implemented by
experimentalists. Moreover, when we notice
\begin{eqnarray}
&&\frac{1}{\pi }\exp [2i\left( p-x\right) \left( q-y\right) ]  \notag \\
&=&\int_{-\infty }^{\infty }\frac{dv}{2\pi }\delta \left( q-y-\frac{v}{2}%
\right) \exp \left\{ i\left( p-x\right) v\right\} ,  \label{21}
\end{eqnarray}%
so the transformation (\ref{1}) is equivalent to%
\begin{eqnarray}
h(p,q) &\rightarrow &\iint_{-\infty }^{\infty }\frac{dpdq}{\pi }e^{2i\left(
p-x\right) \left( q-y\right) }h(p,q)  \notag \\
&=&\iint_{-\infty }^{\infty }dpdq\int_{-\infty }^{\infty }\frac{dv}{2\pi }%
\delta \left( q-y-\frac{v}{2}\right) e^{i\left( p-x\right) v}h(p,q)  \notag
\\
&=&\iint_{-\infty }^{\infty }\frac{dpdq}{2\pi }h(p+x,y+\frac{q}{2})e^{ipq}.
\label{22}
\end{eqnarray}%
For example, using (\ref{7}) and (\ref{22}) we have
\begin{eqnarray}
\Delta (p,q) &\rightarrow &\iint_{-\infty }^{\infty }\frac{dpdq}{2\pi }%
\Delta (p+x,y+\frac{q}{2})e^{ipq}  \notag \\
&=&\iint_{-\infty }^{\infty }\frac{dpdq}{4\pi ^{2}}\int_{-\infty }^{\infty
}due^{-i\left( p+x\right) u}  \notag \\
&&\times \left\vert y+\frac{q}{2}-\frac{u}{2}\right\rangle \left\langle y+%
\frac{q}{2}+\frac{u}{2}\right\vert e^{ipq}  \notag \\
&=&\int_{-\infty }^{\infty }\frac{dq}{2\pi }\int_{-\infty }^{\infty
}due^{-ixu}\delta \left( q-u\right)   \notag \\
&&\times \left\vert y+\frac{q}{2}-\frac{u}{2}\right\rangle \left\langle y+%
\frac{q}{2}+\frac{u}{2}\right\vert   \notag \\
&=&\int_{-\infty }^{\infty }\frac{du}{2\pi }e^{-ixu}\left\vert
y\right\rangle \left\langle y+u\right\vert =\left\vert y\right\rangle
\left\langle y\right\vert \int_{-\infty }^{\infty }\frac{du}{2\pi }%
e^{iu\left( P-u\right) }  \notag \\
&=&\delta \left( y-Q\right) \delta \left( x-P\right) ,  \label{23}
\end{eqnarray}%
so%
\begin{equation}
\frac{1}{\pi }\iint \mathtt{d}p^{\prime }\mathtt{d}q^{\prime }\Delta \left(
q^{\prime },p^{\prime }\right) e^{2\mathtt{i}\left( p-p^{\prime }\right)
\left( q-q^{\prime }\right) }=\delta \left( q-Q\right) \delta \left(
p-P\right) ,  \label{24}
\end{equation}%
thus this new transformation can convert the Wigner function of a density
operator $\rho $, $W_{\psi }(p,q)\equiv \mathtt{Tr}\left[ \rho \Delta (p,q)%
\right] ,$ to%
\begin{eqnarray}
&&\iint_{-\infty }^{\infty }\frac{dp^{\prime }dq^{\prime }}{\pi }\mathtt{Tr}%
\left[ \rho \Delta (p^{\prime },q^{\prime })\right] e^{2\mathtt{i}\left(
p-p^{\prime }\right) \left( q-q^{\prime }\right) }  \notag \\
&=&\mathtt{Tr}\left[ \rho \delta \left( q-Q\right) \delta \left( p-P\right) %
\right]   \notag \\
&=&\int \frac{dudv}{4\pi ^{2}}\mathtt{Tr}\left[ \rho e^{i\left( q-Q\right)
u}e^{i\left( p-P\right) v}\right] ,  \label{25}
\end{eqnarray}%
we may define $\mathtt{Tr}\left[ \rho e^{i\left( q-Q\right) u}e^{i\left(
p-P\right) v}\right] $ as the $Q-P$ characteristic function. Similarly,
\begin{eqnarray}
&&\iint_{-\infty }^{\infty }\frac{dp^{\prime }dq^{\prime }}{\pi }\mathtt{Tr}%
\left[ \rho \Delta (p^{\prime },q^{\prime })\right] e^{-2\mathtt{i}\left(
p-p^{\prime }\right) \left( q-q^{\prime }\right) }  \notag \\
&=&\mathtt{Tr}\left[ \rho \delta \left( p-P\right) \delta \left( q-Q\right) %
\right]   \notag \\
&=&\int \frac{dudv}{4\pi ^{2}}\mathtt{Tr}\left[ \rho e^{i\left( p-P\right)
v}e^{i\left( q-Q\right) u}\right]   \label{26}
\end{eqnarray}%
we name $\mathtt{Tr}\left[ \rho e^{i\left( p-P\right) v}e^{i\left(
q-Q\right) u}\right] $ as the $P-Q$ characteristic function.

In summary, we have found a new integration transformation which can convert
chirplet function to FrFT kernel, this new transformation is worth paying
attention because it is invertible and obeys Parseval theorem. Under this
transformation the relationship between a phase space function and its
Weyl-Wigner quantum correspondence operator is revealed.

\textbf{ACKNOWLEDGEMENT:} Work supported by the National Natural Science
Foundation of China under grants: 10775097 and 10874174.

\bigskip

\bigskip

\bigskip

\bigskip

\bigskip

\bigskip

\bigskip

\end{document}